\documentclass[letterpaper, conference]{IEEEtran}

\usepackage[backend=bibtex,
     bibstyle=ieee,
     sortcites=true,
     maxnames=2,
     maxbibnames=12,
     minnames=1,
     mincrossrefs=99,
     uniquelist=false,
     defernumbers=true
]{biblatex}

\makeatletter
\newcounter{IEEE@bibentries}
\renewcommand\IEEEtriggeratref[1]{%
 \renewbibmacro{finentry}{%
   \stepcounter{IEEE@bibentries}%
   \ifthenelse{\equal{\value{IEEE@bibentries}}{#1}}
   {\finentry\@IEEEtriggercmd}
   {\finentry}%
 }%
}
\makeatother

\usepackage{booktabs}
\usepackage{tabularx}
\usepackage{csquotes}

\usepackage{tikz}
\usepackage{tikzpeople}
\usetikzlibrary{trees}
\usetikzlibrary{positioning}
\usetikzlibrary{arrows.meta,arrows}
\usetikzlibrary{decorations.pathreplacing}
\usepackage[edges]{forest}
\forestset{qtree/.style={for tree={parent anchor=south,
           child anchor=north,align=center,inner sep=0pt}}}

\usepackage{pgfplots}
\usetikzlibrary{
        pgfplots.dateplot,
        pgfplots.groupplots,
        external,
    }
\usepackage{pgfplotstable}
\pgfplotstableread[col sep=comma]{con_versions}\datatable
\pgfplotstableread[col sep=comma]{con_protocols}\datatablei

\pgfdeclarelayer{bg}
\pgfsetlayers{bg,main}

\usepackage{cleveref}
\crefname{figure}{\figurename}{Tables}
\crefname{table}{\tablename}{Figures}

\usepackage{xspace}
\newcommand*{\eg}{e.g.\@\xspace}
\newcommand*{\ie}{i.e.\@\xspace}

\definecolor{sron0}{HTML}{332288}
\definecolor{sron1}{HTML}{88CCEE}
\definecolor{sron2}{HTML}{117733}
\definecolor{sron3}{HTML}{DDCC77}
\definecolor{sron4}{HTML}{CC6677}
\definecolor{sron5}{HTML}{AA4499}

\begin{document}

\author{\IEEEauthorblockN{Erik Daniel}
\IEEEauthorblockA{\textit{Distributed Security Infrastructures} \\
\textit{Technische Universit\"at Berlin}\\
erik.daniel@tu-berlin.de}
\and
\IEEEauthorblockN{Florian Tschorsch}
\IEEEauthorblockA{\textit{Distributed Security Infrastructures} \\
\textit{Technische Universit\"at Berlin}\\
florian.tschorsch@tu-berlin.de}
}

\title{Passively Measuring IPFS Churn and Network Size}

\maketitle

\begin{abstract}
The InterPlanetary File System~(IPFS) is a popular
decentralized peer-to-peer network for exchanging data.
While there are many use cases for IPFS,
the success of these use cases depends on the network.
In this paper, we provide a passive measurement study
of the IPFS network, investigating peer dynamics and
curiosities of the network.
With the help of our measurement, we estimate
the network size and confirm the results of previous
active measurement studies.
\end{abstract}


\section{Introduction}
In the last years, new peer-to-peer~(P2P) data networks emerged,
providing new opportunities for distributed and decentralized
storage and exchange of files~\cite{daniel2022ipfs}.
A prime example is the InterPlanetary File System~(IPFS)~\cite{benet2014ipfs}.
By combining ideas from BitTorrent, Kademlia, Git, and information-centric networking,
IPFS has become a major supplement for distributed storage solutions.
IPFS is often combined with blockchains serving as
off-chain storage~\cite{hoang2020privacy,chen2017improved,ali2017iot}.
Furthermore, IPFS is the originator and a major user of the \emph{libp2p} library~\cite{gitlibp2p},
which could become a quasi-standard for P2P communication.

In this paper, we take a look at a key component of IPFS: the P2P network.
Our goal is to gain a better understanding of the dynamics of the IPFS network,
with the help of passive measurements.
In particular, the churn rate is an important aspect for identifying
possible protocol and configuration weaknesses.
Furthermore, we investigate changes in peer roles
and are interested in finding an answer to a seemingly simple question:
How large is the IPFS network?
In contrast to active measurement, \eg, crawler~\cite{henningsen2020crawling,gitnebulacrawler}, which might influence
a peer's connection limit, triggering connection trimming,
passive measurements are less intrusive.
Due to the usage of the \emph{libp2p} library,
insights in the IPFS network might be applicable to other P2P projects as well.

By deploying two passive measurement clients (go-ipfs, hydra-booster), 
we record basic information about the network's characteristics.
We perform multiple short-term measurements,
spanning a period of $1\,d$ to $3\,d$ in December 2021.
Concerning the network dynamics, we found that
the default configuration, which defines certain thresholds for connection trimming,
causes a high connection churn rate for DHT-Servers,
resulting in very short-lived connections.
While the client did not have more than $\approx 16k$ simultaneous connections,
the client established connections to $40k$--$65k$ different peer IDs~(PID).
This indicates either a high amount of nodes with changing PIDs,
very volatile nodes, or many one-time users.
The meta data of PIDs remains mostly constant.
Naturally, the client version changes over time
due to the appearance of new versions,
however, we observed up- and downgrades.
A small part of the PIDs changes their role in the network,
by switching from a DHT-Server to a DHT-Client and vice versa.
We also identified some anomalies, \eg, go-ipfs clients
not supporting Bitswap or an Ethereum client.

With our passive measurement study,
we contribute a novel perspective on IPFS peer dynamics
that we utilize to approximate the network size.
To this end, we explore two methods:
distinguishing peers based on meta data
and classifying peers based on their connection behavior.
As a result, we conclude that during our measurement period
the network consisted of roughly $48k$ peers.
Based on the classification the core network of IPFS
has at least a size of $10k$ nodes.
With our results, we can also confirm
the general observations of the overlay network made
by~\citeauthor{henningsen2020mapping}~\cite{henningsen2020mapping}.

The remainder is structured as follows.
In \Cref{sec:relwork}, we give a brief overview of related work.
The measurement method and setup is described in \Cref{sec:setup}.
We present an overview of the measurement results in \Cref{sec:results},
before attempting to estimate the network size in \Cref{sec:netsize}.
\Cref{sec:conclusion} concludes the paper and shows our future research direction.

\section{Related Work}\label{sec:relwork}
There exists some literature evaluating IPFS metrics,
\eg, Bitswap~\cite{de2021accelerating} or in general
I/O performance~\cite{shen2019understanding}.
Here, we focus mainly on paper investigating the IPFS P2P network.
\citeauthor{guidi2021data}~\cite{guidi2021data} investigated the data persistence mechanisms
and made a similar analysis about origin, distribution, and usage of transport protocol of peers.
\citeauthor{prunster2020total}~\cite{prunster2020total} analysed an earlier version
of the IPFS client software identifying the libp2p's connection manager's scoring system 
as a potential vulnerability, allowing a Sybil attack.
The presented attack is mitigated in later versions ($\ge$v0.5).
\citeauthor{balduf2021monitoring}~\cite{balduf2021monitoring} investigate the nature
and information of Bitswap messages, focusing on content privacy.
\citeauthor{henningsen2020mapping}~\cite{henningsen2020mapping}
estimated the size of the network and structure of the overlay network.
The authors first get an impression of the default behavior with passive measurements and
compare the results with results from a developed crawler~\cite{henningsen2020crawling}.  
They further report peer distribution, and uptime of peers.
The measurements of the developed crawler is ongoing and the results
are publicly available\footnote{\url{https://trudi.weizenbaum-institut.de/ipfs_crawler.html} (2022-05-30)}.
Another crawler monitoring peer availability and liveness is
the libp2p DHT \enquote{Nebula Crawler}~\cite{gitnebulacrawler}.

Our passive measurement study is similar to the passive method of~\cite{henningsen2020mapping}.
Through our measurements we can confirm their results and
provide a more recent and closer look at peer dynamics.
Furthermore, due to the passive approach of the measurement,
we provide a different perspective on the IPFS network
compared to different ongoing active measurements.

\section{Measurement Setup}\label{sec:setup}
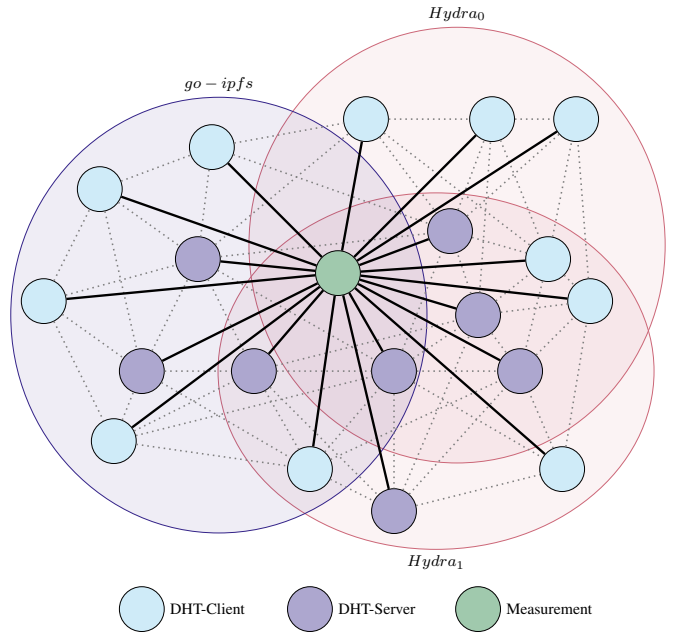
\begin{figure}
\footnotesize
\resizebox{\columnwidth}{!}{
\begin{tikzpicture}
    \tikzstyle{server}=[circle, draw, thin,fill=sron0!40, minimum width=8mm]
    \tikzstyle{client}=[circle, draw, thin,fill=sron1!40, minimum width=8mm]
    \tikzstyle{measurement}=[circle, draw, thin,fill=sron2!40, minimum width=8mm]

    \node [client, label=right:{DHT-Client}] (l0) at (2.00, 1.25) {};
	\node [server, label=right:{DHT-Server}] (l1) at (5.00, 1.25) {};
    \node [measurement, label=right:{Measurement}] (l2) at (8.00, 1.25) {};

    \node [client] (a) at (0.25, 6.75) {};
	\node [client] (b) at (1.25, 8.75) {};
    \node [client] (c) at (1.50, 4.25) {};
	\node [server] (d) at (2.00, 5.50) {};
	\node [client] (e) at (3.25, 9.50) {};
	\node [server] (f) at (3.00, 7.50) {};
	\node [server] (g) at (4.00, 5.50) {};
	\node [client] (h) at (5.00, 3.75) {};
	\node [measurement] (i) at (5.50, 7.25) {};
	\node [client] (j) at (6.00, 10.0) {};
	\node [server] (k) at (6.5, 5.5) {};
	\node [server] (l) at (6.5, 3.00) {};
	\node [server] (m) at (8.00, 6.50) {};
	\node [server] (n) at (7.50, 8.00) {};
	\node [client] (o) at (8.25, 10.0) {};
	\node [client] (p) at (9.50, 3.75) {};
	\node [client] (q) at (9.25, 7.50) {};
	\node [server] (r) at (8.75, 5.50) {};
	\node [client] (s) at (10.0, 6.75) {};
    \node [client] (t) at (9.75, 10.0) {};

 \begin{pgfonlayer}{bg}
    \node [draw=sron4,fill=sron4,fill opacity=0.08,ellipse,
      fit=(i)(k)(t), label={$Hydra_{0}$}] {};
    \node [draw=sron4,fill=sron4,fill opacity=0.08,ellipse,
      fit=(i)(h)(p), label=below:{$Hydra_{1}$}] {};
    \node [draw=sron0,fill=sron0,fill opacity=0.08,ellipse,
      fit=(i)(b)(c), label={$go-ipfs$}] {};
 \end{pgfonlayer}

    \path[thick,dotted,gray] (a) edge (b);
    \path[thick,dotted,gray] (a) edge (f);
    \path[thick,dotted,gray] (a) edge (d);
    \path[thick,dotted,gray] (a) edge (c);
    \path[thick,dotted,gray] (b) edge (f);
    \path[thick,dotted,gray] (b) edge (e);
    \path[thick,dotted,gray] (b) edge (d);
    \path[thick,dotted,gray] (c) edge (d);
    \path[thick,dotted,gray] (c) edge (g);
    \path[thick,dotted,gray] (c) edge (h);
    \path[thick,dotted,gray] (c) edge (k);
    \path[thick,dotted,gray] (d) edge (f);
    \path[thick,dotted,gray] (d) edge (g);
    \path[thick,dotted,gray] (d) edge (h);
    \path[thick,dotted,gray] (e) edge (f);
    \path[thick,dotted,gray] (e) edge (j);
    \path[thick,dotted,gray] (e) edge (n);
    \path[thick,dotted,gray] (f) edge (g);
    \path[thick,dotted,gray] (f) edge (k);
    \path[thick,dotted,gray] (f) edge (j);
    \path[thick,dotted,gray] (f) edge (n);
    \path[thick,dotted,gray] (g) edge (k);
    \path[thick,dotted,gray] (g) edge (h);
    \path[thick,dotted,gray] (g) edge (l);
    \path[thick,dotted,gray] (g) edge (m);
    \path[thick,dotted,gray] (h) edge (k);
    \path[thick,dotted,gray] (h) edge (l);
    \path[thick,dotted,gray] (h) edge (r);
    \path[thick,dotted,gray] (j) edge (n);
    \path[thick,dotted,gray] (j) edge (o);
    \path[thick,dotted,gray] (j) edge (q);
    \path[thick,dotted,gray] (k) edge (n);
    \path[thick,dotted,gray] (k) edge (m);
    \path[thick,dotted,gray] (k) edge (r);
    \path[thick,dotted,gray] (k) edge (l);
    \path[thick,dotted,gray] (k) edge (p);
    \path[thick,dotted,gray] (l) edge (p);
    \path[thick,dotted,gray] (l) edge (r);
    \path[thick,dotted,gray] (l) edge (m);
    \path[thick,dotted,gray] (m) edge (n);
    \path[thick,dotted,gray] (m) edge (q);
    \path[thick,dotted,gray] (m) edge (r);
    \path[thick,dotted,gray] (m) edge (s);
    \path[thick,dotted,gray] (m) edge (o);
    \path[thick,dotted,gray] (n) edge (o);
    \path[thick,dotted,gray] (n) edge (t);
    \path[thick,dotted,gray] (n) edge (q);
    \path[thick,dotted,gray] (o) edge (t);
    \path[thick,dotted,gray] (o) edge (q);
    \path[thick,dotted,gray] (p) edge (r);
    \path[thick,dotted,gray] (p) edge (s);
    \path[thick,dotted,gray] (q) edge (t);
    \path[thick,dotted,gray] (q) edge (s);
    \path[thick,dotted,gray] (q) edge (r);
    \path[thick,dotted,gray] (r) edge (s);
    \path[thick,dotted,gray] (s) edge (t);

    \path[very thick] (a) edge (i);
    \path[very thick] (b) edge (i);
    \path[very thick] (c) edge (i);
    \path[very thick] (d) edge (i);
    \path[very thick] (e) edge (i);
    \path[very thick] (f) edge (i);
    \path[very thick] (g) edge (i);
    \path[very thick] (h) edge (i);
    \path[very thick] (j) edge (i);
    \path[very thick] (k) edge (i);
    \path[very thick] (l) edge (i);
    \path[very thick] (m) edge (i);
    \path[very thick] (n) edge (i);
    \path[very thick] (o) edge (i);
    \path[very thick] (p) edge (i);
    \path[very thick] (q) edge (i);
    \path[very thick] (r) edge (i);
    \path[very thick] (s) edge (i);
    \path[very thick] (t) edge (i);

\end{tikzpicture}
}
  \caption{Illustration of network horizons
  for passive measurements with different measurement clients:
  go-ipfs with a single identity and Hydra with multiple simultaneous identities.}
  \label{fig:netview}
\end{figure}

In general, a network can be measured
actively or passively.
In an active measurement, probing packets
are sent to participants in the network.
While this allows for detailed information
on nodes at almost any time,
it can disrupt normal node behavior.
A passive measurement records data without
creating additional or modified traffic in the network.
Due to the decentralized nature of a P2P network,
the horizon, \ie, the amount of reachable nodes at a time, is limited:
there is no node which knows the exact amount of nodes in the network.
While active nodes aggressively try to maximize their horizon by seeking peers,
the passive nodes' horizon is mainly dependent on its neighbors.
\cref{fig:netview} visualizes the different network perspectives.
In case of IPFS, active measurements like crawlers,
which explore and search the Kademlia-based DHT,
only gain information about peers
actively participating in the DHT (DHT-Server).
A passive measurement can potentially see all peers
independent of their participation in the DHT routing,
which also includes clients (DHT-Client).
The horizon of a \enquote{normal} node, \eg, go-ipfs,
also depends on its position in the DHT,
which determines the priority of
other nodes to maintain or establish a connection to the node.
Special nodes, \eg, Hydras, can establish multiple identities
providing the node with a broader horizon.

Passive measurements are a common method
for gaining an understanding
of peer-to-peer networks~\cite{decker2013propagation, daniel19mapz}.
The basic procedure of a passive measurement is the same for any P2P network.
A node conforming to the protocols behavior is deployed.
This measurement node itself behaves normally,
\ie, answering requests.
At most a node tries to establish and maintain
as many connections as possible,
with the goal to ideally connect
to all nodes in the network.
Internally, the node records data
about the network.

For our measurement, we deployed two different measurement nodes on a VM
hosted in Hesse, Germany by Google Cloud.
The VM had $32\,GB$ RAM, 8 cores, Intel(R) Xeon(R) CPU @ $2.20\,GHz$,
and Ubuntu 21.10. as the operating system~(OS).
During the measurement the VM had a public IPv4 address
and was not externally reachable via IPv6.
Toward the outside, the measurement nodes were a
go-ipfs v0.11.0-dev\footnote{Commit: 0c2f9d5950c4245d89fcaf39dd1baa754587231b}
and a hydra-booster v0.7.4 node.
The clients were mostly deployed simultaneously
on port 4001 for go-ipfs and started by port 3001 for hydra-booster.
Additionally, we deployed another
go-ipfs v0.13.0-dev\footnote{Commit: b2efcf5ce3bba997997962122f85d12500962927}
on a different VM reachable via IPv4 and IPv6.
This VM had $16\,GB$ RAM, 8 cores, Intel(R) Xeon(TM) CPU $3.20\,GHz$,
and Debian 9 as the OS.

We conducted five measurements
with durations between $\approx1\,d$ and $\approx3\,d$.
Between the measurement periods, we adjusted
\emph{LowWater} and \emph{HighWater} values of the connection manager
to reduce connection trimming.
Connections are selectively trimmed to the \emph{LowWater} value,
once the number of simultaneous connections reaches the \emph{HighWater} threshold.
An overview of the measurement periods and the used configuration is provided in~\cref{tab:mperiods}.

\begin{table}
\footnotesize
\centering
\caption{Overview and duration of the measurement periods. Used Versions: go-ipfs v0.11.0-dev/v0.13.0-dev ($P3$), Hydra v0.7.4}
\begin{tabular}{lrrrrr}
\toprule
Period & Measurement Duration & \emph{Low} & \emph{High} & go-ipfs & Hydra \\
\midrule
$P0_{1}$ & 2021-12-03 -- 2021-12-06 &  600 &  900 & Server & -- \\
$P0_{2}$ & 2021-12-03 -- 2021-12-06 & 1.2k & 1.8k & -- & 3 \\
$P1$ & 2021-12-09 -- 2021-12-10 & 2k & 4k & Server & 2 \\
$P2$ & 2021-12-13 -- 2021-12-14 & 18k & 20k & Server & 2 \\
$P3$ & 2022-02-16 -- 2022-02-17 & 18k & 20k & Client & -- \\
$P4$ & 2021-12-10 -- 2021-12-13 & 18k & 20k & Server & -- \\
\bottomrule
\end{tabular}
\label{tab:mperiods}
\end{table}

\subsection{Go-IPFS}
Go-ipfs refers to the
go reference client\footnote{\url{https://github.com/ipfs/go-ipfs} (2022-05-30)}
maintained by Protocol Labs.
The reference client is open source and can be deployed
by anyone to join the IPFS network
or build a private IPFS network.
It supports a wide range of functions to exchange and distribute data.

The client is started utilizing the default configuration,
a random $2048\,bit$ key, and a temporary repository.
Neither key nor repository persisted over different measurement runs.
For the measurements only the \emph{LowWater} and \emph{HighWater} values
of the swarm connection manager were adjusted.

After the initial bootstrap, no further user activity was introduced
such as actively connecting to specific peers or requesting data.
During the deployment every $30\,s$ peer and connection data was recorded and updated.
It recorded for the PID of all known peers in the \emph{Peerstore},
agent version, protocols, and multiaddresses.
Furthermore, changes to the information were recorded with a timestamp.
Additionally, information of all PeerHost connections per connection-id,
\eg, direction, multiaddress, opened, connectedness were checked.
The whole information was exported periodically to a json-file.

The go-ipfs client can be deployed as a DHT-Server or DHT-Client,
determining its participation in the Kademlia-based DHT routing.
Depending on the setting the clients significance in the network changes.
Other nodes rather connect and maintain a connection
to a DHT-Server since the node can answer
routing requests and is detectable via the DHT.
Due to the Kademlia-based DHT peers actively
seek a connection with the measurement node.
The view of a single node, however, is limited
due to its location based on the PID.

\subsection{Hydra-booster}
Hydra-booster, in the following Hydra, refers to a node type designed to
accelerate IPFS content provision and
routing\footnote{\url{https://github.com/libp2p/hydra-booster} (2022-05-30)}.
The node's main purpose is the provision
and collection of DHT records
by deploying multiple \enquote{heads},
which provide basic networking functionality and DHT management,
and the same \enquote{belly} for storing records.
The heads of the Hydra have different PIDs,
allowing the heads to be responsible for different parts of the DHT.

For our measurement, we added two new \emph{PeriodicTasks} to the Hydra,
to update and export measurement data.
The client was configured to record and
update peer data every $1\,min$.
The Hydra recorded each head's known PIDs in the Peerstore,
agent version, protocols, and multiaddresses.
Furthermore, changes to the information were recorded with a timestamp.
Additionally, during a connection and disconnection event
information about the connection, \eg, duration,
timestamp, and direction was logged.
All information was periodically exported to a json-file.

Due to the nature of the Hydra,
the node is a DHT-Server participating in the Kademlia-based DHT routing.
Since every head is located in a different location of the DHT,
the Hydra can cover a broader range of the network,
increasing its range by increasing the number of heads.
As indicated in \cref{fig:netview}, a head's view on the network
is not distinctly different from other heads
and peers might be observed by multiple heads.

\subsection{Measurement Horizon}
\begin{figure}
\footnotesize
\centering
\begin{tikzpicture}

    \begin{axis}[
    ymin=-0.5,
    ymax=4.5,
    no marks,
    hide axis,
    xmin=10000,
    xmax=89000,
    legend columns=4,
    height=5cm,
    width=\columnwidth,
    legend style={at={(0.9,1.1)},anchor=east, draw=none, line width = 1pt,}
    ]
    \addlegendimage{sron1}\addlegendentry{Hydra}
    \addlegendimage{sron2}\addlegendentry{WB Crawler}
    \addlegendimage{sron0}\addlegendentry{go-ipfs}
    \end{axis}

\begin{axis}[
    xbar = 5mm,
    bar width = 4 pt,
    xmin = 10000,
    xmax = 89000,
    xmode=log,
    xtick={10000,15000,25000,40000,65000},
    xticklabels={10k,15k,25k,40k,65k},
    xlabel=\# PID (log scale),
    ylabel=Measurement Period,
    ytick=data,
    yticklabels={P0, P1, P2, P3, P4},
    ymin=-0.5,
    ymax=4.5,
    xbar stacked,
    bar shift =-5pt,
    area style,
    legend style={area legend,at={(0.5,-0.3)},anchor=north,legend columns=-1},
    height=5cm,
    width=\columnwidth,
]
\addplot[sron0,fill=sron0] coordinates{(17347,4) (12899, 3) (13187, 2) (10032,1) (18845,0)};
\addplot[sron0!40, fill=sron0!40] coordinates{(41037,4) (30969, 3) (30746, 2) (5650,1) (47008,0)};

\end{axis}
\begin{axis}[
    xbar = 6mm,
    bar width = 4 pt,
    xmin = 10000,
    xmax = 89000,
    xmode=log,
    ymin=-0.5,
    ymax=4.5,
    ytick=data,
    yticklabels={ ,  ,  ,  ,  },
    xmajorticks=false,
    xbar stacked,
    bar shift=0pt,
    area style,
    legend style={area legend,at={(0.5,-0.3)},anchor=north,legend columns=-1},
    height=5cm,
    width=\columnwidth,
]
\addplot[sron2, fill=sron2] coordinates{(12818,4) (13470, 3) (14383, 2) (20378,1) (13562,0)};
\addplot[sron2!40, fill=sron2!40] coordinates{(2880,4) (3170, 3) (2651, 2) (1559,1) (3400,0)};

\end{axis}
\begin{axis}[
    xbar = 6mm,
    bar width = 4 pt,
    xmin = 10000,
    xmax = 89000,
    xmode=log,
    ymin=-0.5,
    ymax=4.5,
    ytick=data,
    yticklabels={ ,  ,  ,  ,  },
    xmajorticks=false,
    xbar stacked,
    bar shift=5pt,
    area style,
    legend style={area legend,at={(0.5,-0.3)},anchor=north,legend columns=-1},
    height=5cm,
    width=\columnwidth,
]
\addplot[sron1, fill=sron1] coordinates{(22919,4) (16216, 3) (17489, 2) (0,1) (0,0)};
\addplot[sron1!40, fill=sron1!40] coordinates{(65723,4) (39987, 3) (44185, 2) (0,1) (0,0)};

\end{axis}
\end{tikzpicture}
 	\caption{Comparison of the number of peers for active and passive measurements. Showing the number of DHT-Server PIDs (solid) and total PIDs (whole bar) for the passive measurement and a range of min. (solid) and max. (whole bar) PIDs for the active measurement.}
	\label{fig:pid}
\end{figure}
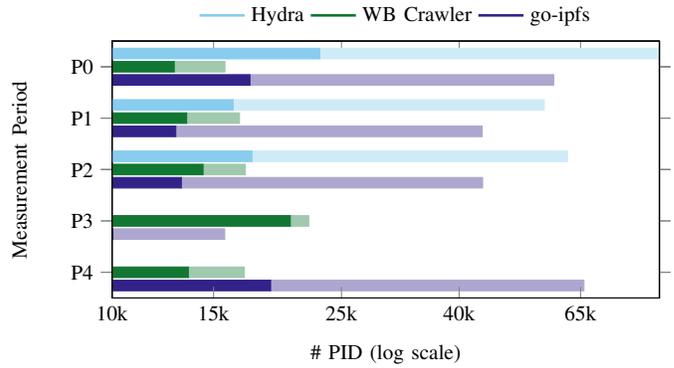

As previously mentioned,
passive measurements can have a limited view on the network.
To determine if our passive measurement covers
a sufficient range of the network,
we compare our measurement results with public crawler
results\footnote{\url{https://trudi.weizenbaum-institut.de/ipfs_crawler.html} (2022-05-30)} (WB Crawler).
\cref{fig:pid} shows the total number of observed PIDs (whole bar)
and the number of identified DHT-Server nodes (solid)
for our passive approaches and an active crawler.
Since crawls are executed every $8\,h$,
the crawler reports varying number of nodes.
We therefore decided to show the results as a range
with the reported min. (solid) and max. (whole bar) values.
The number of PIDs for the Hydra are the union of all heads.
It should be noted that the crawler can only see DHT-Server nodes
and our measurement node can only determine if a node is a DHT-Server,
if protocol information were exchanged between the nodes.

In order to get a complete view of the network,
we should deploy multiple vantage points.
Therefore, we expect to see the least amount of nodes with the go-ipfs client,
more clients with the hydra, and almost all nodes with the crawler.
If we compare the numbers, we can see that for the measurement periods
lasting more than $1\,d$ the passive nodes see more DHT-Server nodes than the crawler.
For the measurement periods of $1\,d$ the Hydra sees
more than the go-ipfs client.
The crawler sees a similar amount of PIDs as the Hydra with two heads.
In $P0$ with three heads, the amount of nodes seen by the Hydra is much higher.
One explanation for the difference are disappearing nodes.
The crawler provides a fresh snapshot of the network
and does not consider results of previous runs.
When peers remove a node's DHT entry,
the crawler cannot gain information about the node,
even though it was previously seen.
Our passive measurement node provides a historic snapshot,
keeping the record of all nodes once seen over time
independent of their activity.

In general, the numbers show a similar range of PIDs.
It therefore seems that one passive measurement node is enough
to reach a reasonable sample of DHT-Server nodes.
Assuming that the crawler covers most of the network
and considering the results shown by the Hydra,
two measurements nodes with strategically placed keys
should be sufficient to cover almost the whole network.
The geolocation of the measurement nodes should not have an influence
on the number of seen nodes
as IPFS does not enforce any geographic clustering of nodes.
However, it is possible that the geolocation of the measurement node
has an influence on the peer dynamics like connection churn.

\section{Peer Dynamics}\label{sec:results}
In this section, we take a closer look at peer dynamics.
We investigate connection churn with
the measurement data from periods $P0$ -- $P3$.
$P4$ which has a longer duration covers observations
of data directly related to the peers.
In the following, we distinguish peers based on their PID.

\subsection{Connection Churn}
To investigate connection dynamics,
we conducted different measurements
with different Low-/HighWater values.
Please note, that due to our measurement setup
in go-ipfs the connection information
is only refreshed every $30\,s$ and the
real values should be slightly smaller than shown.
All connections still active at the end of the measurement
are considered to be closed at that moment
and are included in the statistics.
An overview of the results of the different measurement values can be seen
in~\cref{tab:connections}.
In the statistic, we consider only peers with recorded connection information.
Type \enquote{All} means that the results
represent all connections from all peers,
meaning that some peers provide multiple values.
Type \enquote{Peer} means that the results
are calculated over the average connection duration of each peer,
giving each peer exactly one value.

Overall, the measurements show similar results for the go-ipfs DHT-Server and Hydra,
except for $P0$ where the peer average is much lower compared to the Hydra value.
This exception could be explained by the difference of the default
\emph{LowWater} and \emph{HighWater} values.
For the go-ipfs DHT-Client in $P4$, we see overall short connection durations.

\begin{table}
\footnotesize
\centering
\caption{Connection Statistics for the three measurement periods.}
\begin{tabular}{lllrrr}
\toprule
 & Period & Type & Sum & Avg. & Median \\
\midrule
\multicolumn{6}{l}{\textbf{go-ipfs}}\\
 & $P0$ & All  & 1’285’513 & $196.556\,s$ & $73.732\,s$ \\ 
 & $P0$ & Peer & 55’258 & $695.946\,s$ & $83.008\,s$ \\ 
 & $P1$ & All  & 355'965 & $802.617\,s$ & $130.464\,s$ \\ 
 & $P1$ & Peer & 41'880 & $2'428.966\,s$ & $580.312\,s$ \\ 
 & $P2$ & All  & 285'357 & $3'883.828\,s$ & $85.404\,s$ \\ 
 & $P2$ & Peer & 42'038 & $19'676.930\,s$ & $3'017.252\,s$ \\ 
 & $P3$ & All  & 47'571 & $120.613\,s$ & $75.192\,s$ \\ 
 & $P3$ & Peer & 10'004 & $182.043\,s$ & $72.964\,s$ \\ 
\midrule
\multicolumn{6}{l}{\textbf{Hydra H0}}\\
 & $P0$ & All  & 1'733'511 & $302.257\,s$ & $78.833\,s$ \\ 
 & $P0$ & Peer & 56'465 & $2'445.300\,s$ & $124.226\,s$ \\ 
 & $P1$ & All  & 422'164 & $660.900\,s$ & $76.530\,s$ \\ 
 & $P1$ & Peer & 43'550 & $2'512.923\,s$ & $541.492\,s$ \\ 
 & $P2$ & All  & 416'711 & $2'941.519\,s$ & $65.181\,s$ \\ 
 & $P2$ & Peer & 52'134 & $16'553.299\,s$ & $1'923.119\,s$ \\ 
\midrule
\multicolumn{6}{l}{\textbf{Hydra H1}}\\
 & $P0$ & All  & 1'851'308 & $285.506\,s$ & $78.204\,s$ \\ 
 & $P0$ & Peer & 64'147 & $2'122.097\,s$ & $117.375\,s$ \\ 
 & $P1$ & All  & 538'366 & $524.595\,s$ & $77.110\,s$ \\ 
 & $P1$ & Peer & 43'810 & $2'099.077\,s$ & $439.847\,s$ \\ 
 & $P2$ & All  & 408'621 & $3'003.313\,s$ & $65.339\,s$ \\ 
 & $P2$ & Peer & 48'889 & $18'049.269\,s$ & $2'365.113\,s$ \\ 
\midrule
\multicolumn{6}{l}{\textbf{Hydra H2}}\\
 & $P0$ & All  & 1'890'556 & $280.438\,s$ & $79.585\,s$ \\ 
 & $P0$ & Peer & 63'981 & $1'883.970\,s$ & $113.643\,s$ \\ 
\bottomrule
\end{tabular}
\label{tab:connections}
\end{table}

We observe rather low connection durations lasting
in general only a few minutes up to $1\,h$.
The lower average value of all connections indicates peers
initiating many short lasting connections.
An example for such peers are crawlers,
which connect to a node query their information, \eg, DHT buckets and
then close the connection.

The increase of the connection duration
between measurement periods indicates, however,
that more connections are closed due to connection trimming
than due to nodes leaving the network.
Taking a closer look at the connection type,
we observe vastly more inbound than outbound connections.
In all periods, the duration of inbound connections
is longer than the duration of outbound connections,
which confirms the assumption that the connections are
mainly closed due to connection trimming.
With higher threshold values, the measurement node
no longer trims its own connection,
however, the connection still gets trimmed by the other nodes
which most likely use the default threshold values.

\subsection{Meta Data}
To increase the versatility and for a better understanding
of the roles of specific nodes, nodes reveal information about themselves,
\eg, agent version, supported protocols, and reachable multiaddresses.
This information can be used to,
\eg, identify DHT-Server nodes.
It can also be used to estimate node behavior and identify anomalies.

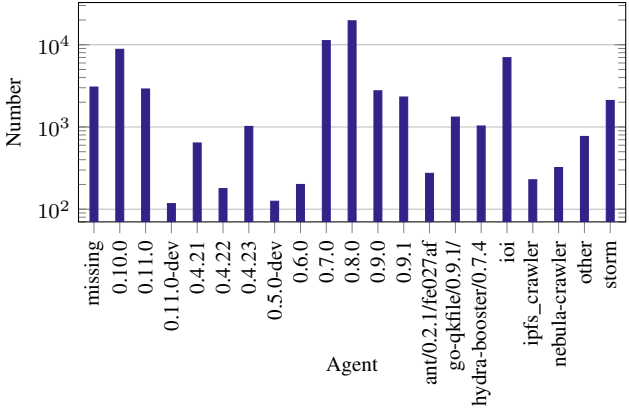
\begin{figure}
\begin{tikzpicture}
  \footnotesize
  \begin{axis}[
  ybar=0pt,
  ylabel=Number,
  xlabel=Agent,
  x label style={at={(axis description cs:0.5,-0.4)}},
  y label style={at={(axis description cs:0.05,0.5)}},
  xtick=data,
  xticklabels from table={\datatable}{agent},
  xtick pos=left,
  ytick pos=both,
  x tick label style={rotate=90},
  bar width=3.0,
  enlarge x limits=0.03,
  no marks,
  height=4.5cm,
  width=\columnwidth,
  ymajorgrids,
  ymode=log,
      ]
  \addplot+[sron0,fill=sron0] table [x expr=\coordindex,y=share,col sep=comma]{con_versions};
  \end{axis}
  \end{tikzpicture}
  \caption{Occurrences of the different agent version strings.}
  \label{fig:versions}
\end{figure}

\begin{figure}
\begin{tikzpicture}
  \footnotesize
  \begin{axis}[
  ybar=0pt,
  ylabel=Number,
  xlabel=Protocol,
  x label style={at={(axis description cs:0.5,-0.7)}},
  y label style={at={(axis description cs:0.05,0.5)}},
  xtick=data,
  xticklabels from table={\datatablei}{agent},
  xtick pos=left,
  ytick pos=both,
  ytick={0,10,100,1000,10000,40000},
  x tick label style={rotate=90},
  bar width=3.0,
  enlarge x limits=0.03,
  no marks,
  height=4.5cm,
  width=\columnwidth,
  legend cell align={right},
  ymode=log,
  ymajorgrids,
      ]
  \addplot+[sron0,fill=sron0] table [x expr=\coordindex,y=share,col sep=comma]{con_protocols};
  \end{axis}
  \end{tikzpicture}
  \caption{Occurrences of the different supported protocols.}
  \label{fig:protocols}
\end{figure}
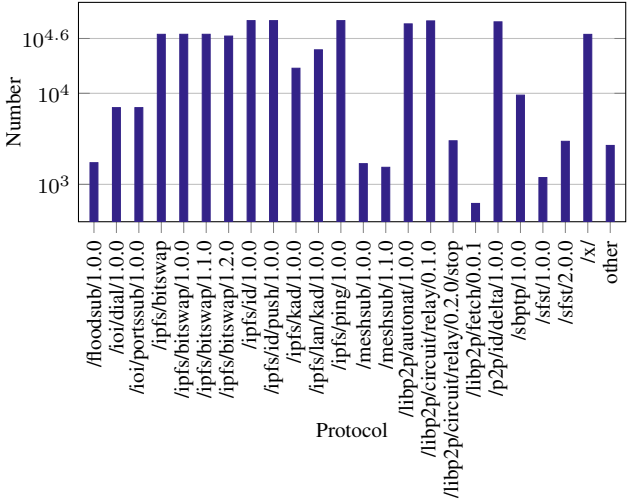

From the data set, we observe overall 323
different agent strings and
101 different supported protocols.
\cref{fig:versions} and \cref{fig:protocols} show the occurrences
of the different agents and protocols.
For a better overview, agents used by 100 or less and protocols
supported by 300 or less are grouped as other.
The go-ipfs versions are grouped by their version number.
From the 323 different agent strings 263 are
different versions of go-ipfs and 61 are other agent strings.
Overall from the 65'853 known PIDs, 50'254
claim to use some sort of go-ipfs version,
1'028 are Hydra nodes, 586 are some kind of crawler,
10'926 use a different agent, and
from 3'059 no version string was obtained.
One of the agents claimed to be a go-ethereum node.

However, agents are prone to change.
During the three days, we could observe some version changes,
mainly affecting go-ipfs clients.
Surprisingly, we could identify not only upgrades but also downgrades.
An overview of the version changes can be seen in~\cref{tab:version}.
As an upgrade, we define an increase in the version number.
A downgrade is, therefore, a decrease in the version number.
A change is indicated by a change of the commit part of the go-ipfs version string.
A dirty version is a version, containing changes
from the release's main version as indicated by the commit message,
\eg, our used go-ipfs versions are dirty versions.
The version of agents other than go-ipfs
did not change in the observed time period.
Once the agent changed from a non-go-ipfs agent to a go-ipfs agent.

The announced agent can be used as an indicator, for peers' behavior.
This should be further distinguishable by the announced protocol.
Almost all of the clients support the basic IPFS protocols
like \emph{ipfs/id} and \emph{ipfs/ping}.
A few custom protocols are used by only a distinct number of peers.
Surprisingly, only 44'463 support \emph{ipfs/bitswap} when 50'163 use go-ipfs.
There are 7'498 go-ipfs v0.8.0 clients which do not support Bitswap
and instead support \emph{sbptp},
a protocol, which is otherwise only supported by storm nodes.
Furthermore, libp2p/circuit/relay is support by almost all clients.
The \emph{ipfs/kad} protocol, which should indicate an IPFS DHT-Server,
is supported by 18'845 peers.

The combination of agent version and
supported protocol can be used to detect unusual behavior.
In case of curiosities like a go-ipfs agent,
which does not support \emph{ipfs/bitswap},
it reveals information about the configuration or
reveals an attempt to hide potentially malign agents,
\eg, storm nodes, identified to be part of
a botnet~\cite{pripoae2020looking}.

\begin{table}
\footnotesize
\centering
\caption{Overview of go-ipfs version changes.}
\begin{tabular}{lrllr}
\toprule
\multicolumn{2}{l}{Version} & & \multicolumn{2}{l}{Type}\\
\cmidrule{1-2}
\cmidrule{4-5}
Upgrade & 218 & & main--main & 291\\
Downgrade & 107 & & dirty--main & 9 \\
Change & 205 & & main--dirty & 5 \\
 & & & dirty--dirty & 225 \\
\bottomrule
\end{tabular}
\label{tab:version}
\end{table}

During normal operation, we do not expect a change
in the announced supported protocols, except
when the used client changes.
In general, only a few nodes changed their supported protocols.
However, we observed many changes concerning \emph{/ipfs/kad/1.0.0}
and \emph{/libp2p/autonat/1.0.0}.
2'481 peers changed in sum 68'396 times their support announcement
of the \emph{/ipfs/kad/1.0.0} protocol switching
their roles from a DHT-Server to DHT-Client.
3'603 peers changed in sum 86'651 times their support announcement
of \emph{/libp2p/autonat/1.0.0}.
While autonat is not a crucial feature, switching roles
from a DHT-Server to a DHT-Client could have
a negative influence on the network,
\eg, due to many control messages.

While the meta data is useful for the operation of the network,
it shows potential for misuse.
Due to the almost homogeneous distribution of agent versions
and supported protocols, exotic combinations or exotic names
could be used to re-identify or track specific peers independent of their PID.
Furthermore, the meta data seems to be rather constant,
which means that this information could be further utilized for peer identification.
However, long term observations are necessary to confirm or deny this assumption.

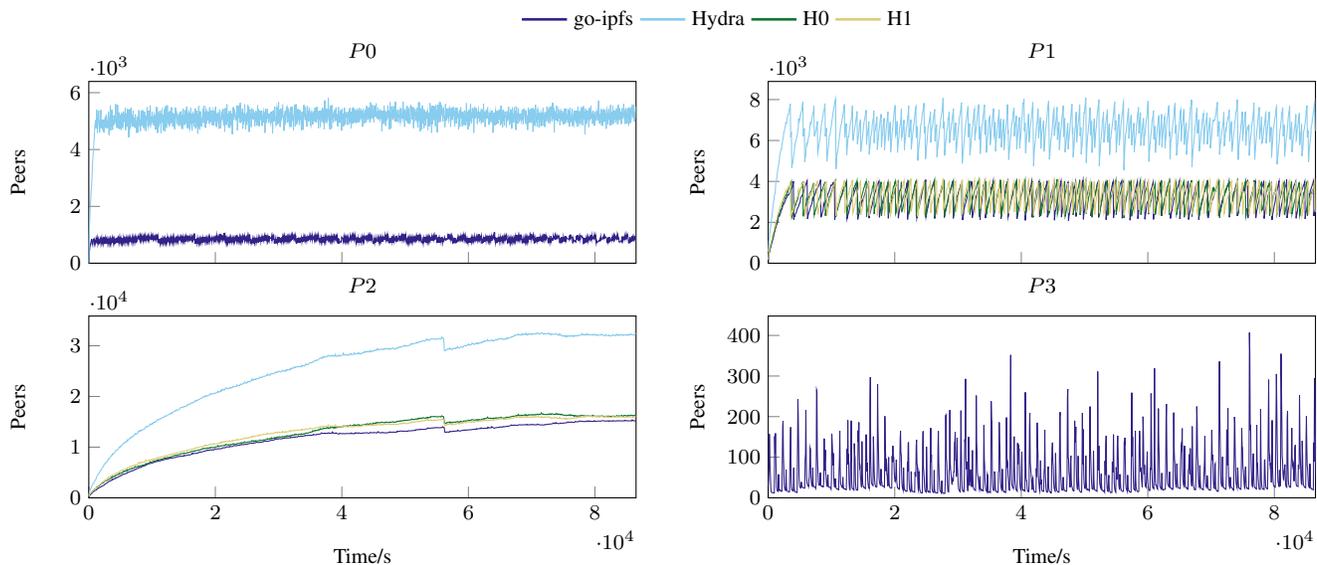
\begin{figure*}
 \footnotesize
 \begin{tikzpicture}
    \begin{axis}[
    ymin=0,
    ymax=1,
    no marks,
    hide axis,
    xmin=0,
    xmax=50,
    legend columns=4,
    width=4in,
    height=1.9in,
    legend style={at={(1.3,1)},anchor=east, draw=none, line width = 1pt,}
    ]
    \addlegendimage{sron0}\addlegendentry{go-ipfs}
    \addlegendimage{sron1}\addlegendentry{Hydra}
    \addlegendimage{sron2}\addlegendentry{H0}
    \addlegendimage{sron3}\addlegendentry{H1}
    \end{axis}

    \begin{groupplot} [
    group style={group name=spr,group size=2 by 2, x descriptions at=edge bottom, horizontal sep=50pt, vertical sep=20pt,},
     xlabel=Time/s,
     ylabel=Peers,
     xmin=0,
     xmax=86500,
     ymin=0,
     height=4cm,
     width=\columnwidth,
     xticklabel style={anchor=near xticklabel},
     xtick pos=left,
     ytick pos=left,
     x label style={at={(axis description cs:0.5,-0.02)}},
     y label style={at={(axis description cs:0.04,0.5)}},
     ]
    \nextgroupplot[title=$P0$,scaled y ticks=base 10:-3]
        \addplot+[sron0,no marks, filter discard warning=false] table [col sep=comma,x=sec,y=con] {p1g};
        \addplot+[sron1,no marks, each nth point=10, filter discard warning=false] table [col sep=comma,x=sec,y=con] {p1hy};
    \nextgroupplot[title=$P1$, scaled y ticks=base 10:-3,]
    \addplot+[sron0,no marks, filter discard warning=false] table [col sep=comma,x=sec,y=con] {p2go};
    \addplot+[sron1,no marks, filter discard warning=false] table [col sep=comma,x=sec,y=hydra] {p2hy};
    \addplot+[sron2,no marks, filter discard warning=false] table [col sep=comma,x=sec,y=h0] {p2hy};
    \addplot+[sron3,no marks, filter discard warning=false] table [col sep=comma,x=sec,y=h1] {p2hy};
    \nextgroupplot[title=$P2$,]
     \addplot+[sron0,no marks, filter discard warning=false] table [col sep=comma,x=sec,y=con] {p3go};
     \addplot+[sron1,no marks, filter discard warning=false] table [col sep=comma,x=sec,y=hydra] {p3hy};
     \addplot+[sron2,no marks, filter discard warning=false] table [col sep=comma,x=sec,y=h0] {p3hy};
     \addplot+[sron3,no marks, filter discard warning=false] table [col sep=comma,x=sec,y=h1] {p3hy};
    \nextgroupplot[title=$P3$,]
     \addplot+[sron0,no marks, filter discard warning=false] table [col sep=comma,x=sec,y=con] {p4go};
    \end{groupplot}
 \end{tikzpicture}
 	\caption{Comparison of the trend of the total simultaneous peer connections.}
	\label{fig:connections}
\end{figure*}

\begin{figure}
 \footnotesize
  \begin{tikzpicture}
    \begin{axis}[
    ymin=0,
    ymax=1,
    no marks,
    hide axis,
    xmin=0,
    xmax=50,
    legend columns=4,
    width=4in,
    height=1.9in,
    legend style={at={(0.7,1)},anchor=east, draw=none, line width = 1pt,}
    ]
    \addlegendimage{sron0}\addlegendentry{all}
    \addlegendimage{sron1}\addlegendentry{$\ge3\,d$ not connected}
    \end{axis}

    \begin{axis}[
    xlabel=Time/h,
    ylabel=\# PIDs,
    xmin=0,
    ymin=0,
    height=4cm,
    width=\columnwidth,
    xtick pos=left,
    ytick pos=both,
    y label style={at={(axis description cs:0.04,0.5)}},
    xmajorgrids,
    ymajorgrids,
    ]
    \addplot+[sron0, no marks, line width=1pt, filter discard warning=false, each nth point=3] table [col sep=comma,x=sec,y=all] {pids};
    \addplot+[sron1, no marks, line width=1pt, filter discard warning=false, each nth point=3] table [col sep=comma,x=sec,y=con] {pids};
    \end{axis}
  \end{tikzpicture}
	\caption{Number of PIDS over time.}
	\label{fig:pids}
\end{figure}
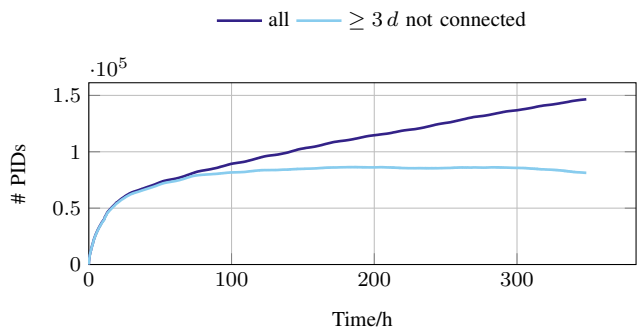

\begin{figure*}
 \footnotesize
 \begin{tikzpicture}
    \begin{axis}[
    ymin=0,
    ymax=1,
    no marks,
    hide axis,
    xmin=0,
    xmax=50,
    legend columns=4,
    width=4in,
    height=1.9in,
    legend style={at={(1.3,1)},anchor=east, draw=none, line width = 1pt,}
    ]
    \addlegendimage{sron0}\addlegendentry{all}
    \addlegendimage{sron1}\addlegendentry{DHT-Server}
    \addlegendimage{sron2}\addlegendentry{DHT-Client}
    \end{axis}

    \begin{groupplot} [
    group style={group name=spr,group size=2 by 1, horizontal sep=50pt, vertical sep=20pt,},
     xlabel=Connection duration/s,
     ylabel=CDF,
     xmin=1,
     ymin=0,
     height=4cm,
     width=\columnwidth,
     xtick pos=left,
     ytick pos=left,
     y label style={at={(axis description cs:0.05,0.5)}},
     xmode=log,
     xmajorgrids,
     ymajorgrids,
     ]
    \nextgroupplot[title=Connection Duration]
     \addplot+[sron0,no marks, line width=1pt, filter discard warning=false, each nth point=2] table [col sep=comma,x=sec,y=cdfall] {con_dur};
     \addplot+[sron1,no marks, line width=1pt, filter discard warning=false, each nth point=2] table [col sep=comma,x=sec,y=cdfkad] {con_dur};
     \addplot+[sron2,no marks, line width=1pt, filter discard warning=false, each nth point=2] table [col sep=comma,x=sec,y=cdfnon-kad] {con_dur};
    \nextgroupplot[title=Distribution \# Connections, xlabel=\# Connections,]
     \addplot+[sron0,no marks, line width=1pt, filter discard warning=false] table [col sep=comma,x=con,y=cdfall] {con_dist};
     \addplot+[sron1,no marks, line width=1pt, filter discard warning=false] table [col sep=comma,x=con,y=cdfkad] {con_dist};
     \addplot+[sron2, no marks, line width=1pt, filter discard warning=false] table [col sep=comma,x=con,y=cdfnon-kad] {con_dist};
    \end{groupplot}
 \end{tikzpicture}
 	\caption{CDF of the maximum connection duration of all PIDs, grouped into $30\,s$ intervals and the CDF of the number of connections.}
	\label{fig:durdist}
\end{figure*}
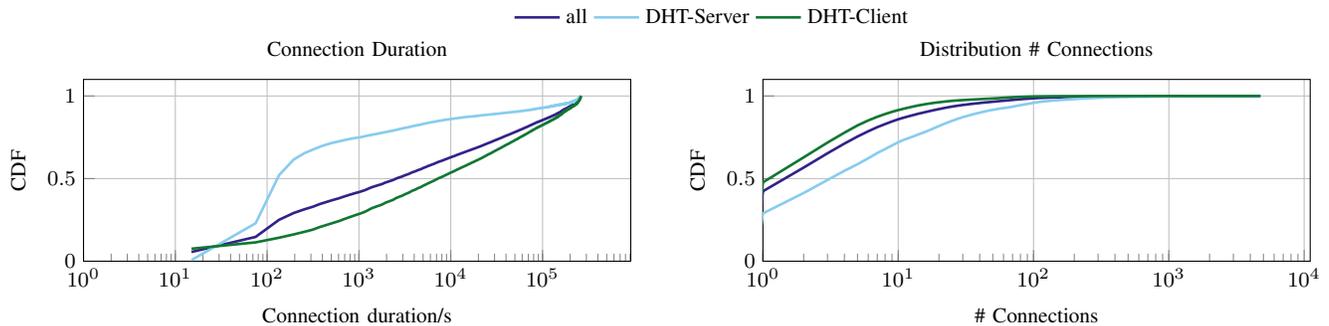

\section{Network size}\label{sec:netsize}
Previously, we differentiated peers by their PID.
Therefore, the network size would be equal to the number of PIDs.

Looking at the number of PIDs in~\cref{tab:connections},
we see between $40k$--$65k$ peers.
Since the position in the network and
the day of the measurements vary,
it is plausible that the number of peers vary as well.
However, looking at the number of simultaneous connections
provides a different perspective on these numbers.
A comparison of the number of connections over time
for the different periods can be seen in~\cref{fig:connections}.
For better comparison, the figure shows only the connections of the first $24\,h$.
In case of $P0$ and $P1$, we can directly see our nodes' connection trimming
due to the configured threshold values.
For $P2$, we can see around 15k--16k simultaneous connection,
which does not even reach the \emph{LowWater} threshold.
$P4$ probably shows other nodes' connection trimming,
since even peak values are below the default go-ipfs \emph{LowWater} value.
The passive DHT-Client node, which does not even provide any files
is a prime candidate for connection trimming.
Naturally, not every peer needs to establish a connection to our measurement node.
Investigating the number of PIDs over a longer time, however, shows a certain trend.
In \cref{fig:pids}, we can see the number of PIDs over time and
the number of PIDs that were more than three days
disconnected from the measurement node and never returned.
To this end, we conducted an additional measurement
of approximately $14\,d$ (from 2022-03-29 to 2022-04-12).
The measurement shows a continuous grow in the number of seen PIDs
and a plateau of connected PIDs.
The difference between simultaneous connections, connected peers, and
the amount of seen PIDs over time are
indicators for a possible misinterpretation of one peer as one PID.

In general, one participant can use multiple PIDs, \eg,
to use a different outer profile, different IPNS entries,
or increasing the privacy against nodes
monitoring the Bitswap traffic~\cite{balduf2021monitoring}.
The difference between simultaneous connections and known PIDs
in our measurement would suggest that every peer has around two PIDs.

To determine the size of the network, it is therefore necessary
to group PIDs or find a different method to distinguish peers.
In the following, we discuss two approaches to estimate the network size:
based on the multiaddress and
based on the connection time and number of connections.

\subsection{Multiaddress}
A simple approach to group PIDs
is the usage of the connected multiaddress, especially, the IP address part.
PIDs establishing a connection
from the same IP address belong to the same group.
In case of the data of $P4$ with 65'853 PIDs,
we had a connection with 62'204 PIDs,
communicating from 56'536 different IP addresses.
Grouping the PIDs by connected IP address results in 47'516 different groups,
with 44'301 groups consisting of only one PID and
overall 40'193 PIDs with unique IP addresses.

This method can detect non-persistent or rotating PIDs to some degree.
For example, we observed one IP address with 2'156 PIDs,
where all PIDs have the same agent version and support the same protocols.
However, the amount of groups are still three times the amount
of simultaneous connections.
Additionally, this method has some serious flaws:
Hydra nodes operate a multitude of heads not necessarily
deployed on different IP addresses.
In our data set, the 1'028 Hydra nodes 1'026 operate
from 11 IP addresses 9 with 100, one with 98, and one with 28 nodes.
The last two are run on one IP address with two go-ipfs nodes.
These are most likely simultaneously active peers but would be grouped to 12 peers.
Other problems consist in non-persisting PIDs in combination with NAT, NAT in general,
smaller Cloud provider sharing IP addresses,
or one-time users.

\subsection{Connection Time}
In addition to distinguish peers based on IP addresses,
we also make an attempt to classify peers
based on connection information.
To this end, we will consider the number of connections
established by a peer
and duration of a connection with a PID.
This goes beyond differentiating between
clients and non-clients~\cite{henningsen2020mapping}
and is able to identify the core network.

To see if there are certain trends,
we use the data of $P4$.
The left side of~\cref{fig:durdist} shows the cumulative distribution function~(CDF) for the
maximum duration per PID for DHT-Clients, DHT-Server, and all PIDs.
Around 53\,\% are connected less than $1\,h$ and
around 16\,\% of the nodes maintained a connection
longer than $24\,h$.
The trend for shorter durations of DHT-Server nodes,
might be due to the connection trimming as also experienced by our node.
On the right side of~\cref{fig:durdist}, we can see the CDF for the
number of connections a PID had with the measurement node.
The figure shows that only around 10\,\% have more than 15 connections
and around 50\,\% have one connection.
From the results of \cref{fig:durdist},
we can conclude that most PIDs connected only a few times
and a small amount of nodes maintained the connection for a long time.

In order to classify peers based on connection information,
we define four classes: heavy, normal, light, and one-time.
We consider heavy peers to be stable and constantly active
\ie, more than a day.
Normal peers have comparably shorter
but still somewhat long connection durations,
\eg, more than two hours but less than a day.
Light peers have many short connections
\ie, repeatedly connecting to the networking,
which summarizes recurring, experimental, faulty,
or malicious peers.
Lastly, one-time peers connect once or twice
to the network in a short time frame
for a short period of time (max. $2\,h$).

\begin{table}
\footnotesize
\centering
\caption{Classification of peers in the $P4$ data set.}
\begin{tabular}{lrrrr}
\toprule
Class & Time & \# Conn. & Peers & DHT-Server \\
\midrule
Heavy & $>24\,h$ & -- & 10'540 & 1'449 \\
Normal & $>\phantom{2}2\,h$ & -- & 15'895 & 1'420 \\
Light & $\leq \phantom{2}2\,h$ & $\geq3$ & 16'880 & 9'755 \\
One-time & $<\phantom{2}2\,h$ & $<3$ & 18'889 & 6'108 \\
\bottomrule
\end{tabular}
\label{tab:classes}
\end{table}

\cref{tab:classes} shows the classification of peers.
We observed $\approx1.5k$ heavy DHT-Server nodes,
yielding $\approx9k$ heavy DHT-client nodes.
Since the DHT clients do not participate in DHT routing,
we can also refer to them as the core user base.
The number of heavy DHT-Server nodes seems rather low,
which is not surprising as it represents a subset of all core nodes only.
That is, some light and one-time DHT-Server nodes
might be core nodes as well.
This misclassification of nodes is a weakness of the passive measurement approach,
because we can only see connection churn and not node churn.
Connection churn can be triggered by connection trimming,
due to new nodes joining the network, or
new connections caused by file exchange and search.
While the number of core nodes might be higher,
it is, however, unlikely that the number is lower.

\section{Conclusion and Future Work}\label{sec:conclusion}
In this paper, we presented a passive measurement study
of the IPFS network, revealing a different perspective
of the network compared to previous active measurements.
While in many other P2P networks
connection churn can be expected to be approximately equal to node churn,
it is different in IPFS:
We identify a rather high connection churn.
Instead of nodes joining and leaving the network,
we believe that the reason for the high connection churn
is IPFS's connection trimming mechanism.
Due to these results, we recommend to investigate
the default threshold values for DHT-Server nodes further
and possibly adjust it to a higher value.

Moreover, we show the complexity of determining the network size.
To this end, we presented two methods for estimating network size
and conclude that both methods cannot capture the full complexity.
Additionally, the short measurement period,
cannot capture long-term behavior.
In the future, we will investigate,
if it is possible to improve our method to distinguish peers
by utilizing a wide range of peer meta data,
\eg, latency, agent version, and history of announced multiaddresses.
However, the possibility of a clear distinction of every peer
can be cause of concern considering the privacy of users.

\IEEEtriggeratref{12}
\printbibliography[heading=bibintoc]

\end{document}